\documentclass[amssymb,prl,twocolumn,longbibliography,notitlepage,superscriptaddress]{revtex4-1}

\newcommand{\kets}[1]{| #1 \rangle}
\newcommand{\bras}[1]{\langle #1 |}

\usepackage{float}
\usepackage{color}

\usepackage{amsmath}                          % better math control.
\usepackage{amsfonts}                         % extended math fonts.
\usepackage{amssymb}                          % names amsfont symbols.
\usepackage{amsthm}                           % theorem environments
\usepackage{mathdots}                         % for ddots going upwards

\usepackage{graphicx}                         % ps, pdf figs.
\usepackage{subfigure}                        % for subfigures
\usepackage{overpic}

\begin{document}

%\title{Digital quantum simulation of finite-temperature quantum system assisted with a cavity mode}
\title{Continuous-variable assisted thermal quantum simulation}

\author{Dan-Bo Zhang}
%\email{dbzhang@m.scnu.edu.cn}
\affiliation{Guangdong Provincial Key Laboratory of Quantum Engineering and Quantum Materials, GPETR Center for Quantum Precision Measurement and SPTE, South China Normal University, Guangzhou 510006, China}
\affiliation{Frontier Research Institute for Physics,
South China Normal University, Guangzhou 510006, China}

\author{Guo-Qing Zhang}
\affiliation{Guangdong Provincial Key Laboratory of Quantum Engineering and Quantum Materials, GPETR Center for Quantum Precision Measurement and SPTE, South China Normal University, Guangzhou 510006, China}
\affiliation{Frontier Research Institute for Physics,
South China Normal University, Guangzhou 510006, China}

\author{Zheng-Yuan Xue}
\affiliation{Guangdong Provincial Key Laboratory of Quantum Engineering and Quantum Materials, GPETR Center for Quantum Precision Measurement and SPTE, South China Normal University, Guangzhou 510006, China}
\affiliation{Frontier Research Institute for Physics,
South China Normal University, Guangzhou 510006, China}

\author{Shi-Liang Zhu}
\email{slzhu@nju.edu.cn}
%\affiliation{National Laboratory of Solid State Microstructures,	School of Physics, Nanjing University, Nanjing 210093, China}

\affiliation{Guangdong Provincial Key Laboratory of Quantum Engineering and Quantum Materials, GPETR Center for Quantum Precision Measurement and SPTE, South China Normal University, Guangzhou 510006, China}
\affiliation{Frontier Research Institute for Physics,
South China Normal University, Guangzhou 510006, China}

\author{Z. D. Wang}
\email{zwang@hku.hk}
\affiliation{Department of Physics and HKU-UCAS Joint Institute for Theoretical and Computational Physics at Hong Kong, The University of Hong Kong, Pokfulam Road, Hong Kong, China}

\affiliation{Frontier Research Institute for Physics,
South China Normal University, Guangzhou 510006, China}

\begin{abstract}
Simulation of a quantum many-body system at finite temperatures is crucially important  but quite challenging.
Here we present an experimentally feasible  quantum algorithm assisted with continuous-variable for %thermal quantum simulation, %~(TQS),
simulating quantum systems at finite temperatures. Our algorithm has a time complexity scaling polynomially with the inverse temperature  and the desired accuracy. We demonstrate the quantum algorithm by simulating finite temperature phase diagram of the Kitaev model. It is found that the important crossover phase diagram of the Kitaev ring can be accurately simulated by a quantum computer with only a few qubits and thus the algorithm may be readily implemented on current quantum processors. We further propose a protocol implementable with superconducting or trapped ion quantum computers.

%Physical systems exists in nature at finite temperatures and one of the most relevant class of states is arguably thermal state.
%Here we present a continuous-variable~(qumode) assisted quantum algorithm for thermal quantum simulation~(TQS). The quantum algorithm converts thermal %information encoded in a qumode resource state into temperature of a quantm system. An equivalent relation is also revealed that allows to simulate thermal %states at varied temperatures with an fixed easy-to-prepare resource state. As for demonstration, we present TQS of a Kitaev ring in the quantum critical %regime, which can accurately determine the crossover temperature, showing its power for simulating physics governed by an interplay between quantum and
%thermal fluctuations.
%We further give a feasible protocol with superconducting circuit system, which may be readily implemented on near-term quantum processors.
\end{abstract}
% To do list
% the complexity of preparing resource state and the complexity of quantum thermal state. Is there a correspondance?
\maketitle

% to to list:
% rewrite the introduction.
\emph{Introduction.--}
Simulation of quantum many-body systems has been an incentive for building quantum computers~\cite{feynman_82}.
Recent advances of scaling-up quantum processors enable us to simulate steady states and dynamics of a quantum system at zero temprature
to a larger size~\cite{barends_15,bernien_17,kandala_17,zhang_17}, and remarkably, to make a simulation of quantum phase transition to be a reality~\cite{islam_11,zhang_17}.  However, simulation of a quantum many-body system at finite temperatures is even more challenging as the state in a quantum computer is usually a pure quantum state, while  simulating a quantum system at finite temperatures requires to simulate a kind of mixed states, namely quantum thermal~(Gibbs) states. The relevance of thermal states is ubiquitous, and its simulaiton  is not only important for physics itself, such as understanding high-Tc superconductivity~\cite{lee06} , but also can provide quantum speedup for optimization~\cite{van_17}.

Thermal quantum simulation~(TQS) requires a well control of both quantum coherence and temperature, which challenges the current quantum platforms. A quantum algorithm based on quantum phase estimation may involve a large number of auxiliary qubits and complicated quantum circuits, which is not suitable for near-term quantum simulators~\cite{terhal_00,poulin_09,temme_11,riera_12}. Recent hybrid quantum-classical variational algorithms require less quantum resource and are feasible in implementation ~\cite{wu_19,verdon_19,liu_19,chowdhury2020variational,wang_20}, but it should be trained for each Hamiltonian at every temperature and thus is not a general solution. Moreover, it waits for a guarantee of quantum advantages in time complexity.

An alternative way is to use continuous variables ~(CV and also called qumode) for encoding and processing high-density information by exploiting its infinite dimensional Hilbert space~\cite{weedbrook_12,lau_17}. Notably, a hybrid approach of incorporating both qubits and qumodes has been shown to have a potential advantage to make the best-of-both-worlds~\cite{furusawa2011quantum,andersen_15,liu_16,gan_19}. Moreover, the mainstream platforms of quantum computers often have naturally existing continuous variables, such as motional modes of trapped ions~\cite{leibfried_03,monroe_13} and cavity modes of superconducting circuits~\cite{wallraff_04,paik_11,devoret_13}, making a hybrid variable approach physical realizable. This may enable us to design hybrid-variable quantum algorithms for TQS with both quantum advantage and feasible physical implementation.

In this paper, we present a quantum algorithm for thermal quantum simulation assisted with an auxiliary qumode, which has a time complexity scaling polynomially with the inverse temperature $\beta$ and the desired accuracy $\epsilon$. The algorithm converts thermal information, encoded in the CV resource state parameterized with $\beta$, into temperature of the quantum system. Moreover, by revealing an equivalence relation of the quantum algorithm, we further propose adaptive TQS, allowing
TQS at varied $\beta$ with a proper-chosen resource  state, which enables flexibility of algorithm design in practice.  To show the power of adaptive TQS, we consider thermal states of two textbook models, the quantum Ising model \cite{Sachdev} and Kitaev ring \cite{kitaev2001unpaired}, in the quantum critical regime, and demonstrate our algorithm can accurately determine the crossover temperature. This indicates an interplay between quantum and thermal fluctuations, which underlines the quantum criticality  at finite temperatures, can be faithfully captured. An intriguing result here is that the important crossover phase diagram of the Kitaev model in periodic condition (Kitaev ring) can be accurately simulated by a quantum computer with only a few qubits and thus the algorithm may be readily implemented on current quantum processors. We also propose an experimental protocol implementable with a superconducting or trapped ion quantum computer. Our work opens a new avenue for simulating finite temperature quantum systems by exploiting the power of continuous variables.

\emph{Thermal quantum simulation.}
Consider a quantum system described by a Hamiltonian $H$, which can be mapped to a quantum computer with $N$ qubits. The energies and eigenstates of the Hamiltonian governed by the Schrodinger equation $H\kets{u_n}=E_n\kets{u_n}$~($n=0,1,...,D-1$,where $D=2^N$). At a finite temperature $T=1/\beta$, the system in equilibrium is in a quantum thermal state $\rho(\beta)=e^{-\beta H}/Z(\beta)$, where $Z(\beta)=\text{Tr}e^{-\beta H}$ is the partition function. Our goal is to prepare a pure quantum state   $\kets{\psi(\beta)}$ in an enlarger Hilbert space which has the property $\rho(\beta)=\text{Tr}_A |\psi(\beta) \rangle \langle\psi(\beta)|$, where $\text{Tr}_A$ denotes the partial trace of some ancillary degrees of freedom addressed later. One can verify that, partial trace of the second partite of the thermofield double~(TFD) state defined as
\begin{equation}
\kets{\psi(\beta)}=\sum_n\frac{e^{-\beta E_n/2}}{\sqrt{\mathcal{Z}(\beta)}}\kets{u_n}\otimes\kets{u^*_n}
\end{equation}
 is just $\rho(\beta)$. We propose a quantum process as follows,
\begin{equation}
\kets{\psi(\beta)} =\sqrt{\mathcal{C}}e^{-\beta H/2}\otimes I \kets{\psi(0)},
\end{equation}
where $\mathcal{C}$ is a normalization factor. Here $\kets{\psi(0)}=\frac{1}{\sqrt{D}}\sum_n\kets{n}\otimes\kets{n}$ is
an infinite temperature TFD which  can be written as a product of N copies of Bell state~(see Supplementary Material~(SM) for a derivation), and thus is easy to prepare. The central task is then to construct $e^{-\beta H/2}$.

%which corresponds to
%thermal state at infinite temperature. Notably, $\kets{\psi(0)}$
%$\kets{\psi(0)}=\prod_{i=1}^{N}B_{i,i+N}$, where $B_{i,i+N}=\frac{1}{\sqrt{2}}(\kets{00}+\kets{11})$ is a Bell state.

%Nonunitary from integral of unitaries
%A direct method is to use an nonunitary operator $e^{-\beta H}$ to transform an completely mixed state $\rho_0=I/d$ to the target state $\rho(\beta)$. Constructions of such non-unitaries can refer to an auxiliary system. Earlier quantum algorithms use quantum phase estimation~(QPE) to register eigenvalues $E_i$ to auxiliary qubits and perform conditional rotation based on eigenvalues to get required quantum amplitude $e^{-\beta E_i}$~\cite{poulin_09,bilgin_10,riera_12}. A more efficient way is to express the nonunitary operator as a combination of unitaries, accomplished with auxiliary modes~\cite{chowdhury_17,motta_19,brandão_19}. This methodology of linear-combination-of-unitaries has been developed ~\cite{gui-lu_06,childs_17,van_17,lau_17,gilyen_18,arrazola_19,zhang_19,zhang_20}, where the number of unitaries is limited by the dimension of Hilbert space. The infinite dimensionality of a qumode allows an integral of infinite number of unitaries~\cite{lau_17,arrazola_19,zhang_19}, which has an advantage over auxiliary qubits. Moreover,
%it can leverage well-established math of functional analysis~\cite{van_17,gilyen_18} for algorithmic design.

We introduce an auxiliary qumode  to represent a nonunitary $e^{-\beta H/2}$ as an integral of unitaries~\cite{lau_17,arrazola_19,zhang_19,zhang_20}, which extends the linear-combination-of-unitaries to the case of CV~\cite{gui-lu_06,childs_17,van_17,gilyen_18,arrazola_19}.
Note that
\begin{equation}
e^{-\beta h/2} = \int_{-\infty}^{\infty} dpR(\beta,p)e^{-ihp},~~R(\beta,p)=\frac{2}{\pi}\frac{\beta}{\beta^2+4p^2}
\end{equation}
holds for $h\geq0$. It implies that we shall add a constant to $H$ to guarantee a nonnegative ground state energy in the algorithm.
Then in the basis $\{\kets{u_n}\}$ that $H$ is diagonal, one can verify that
\begin{eqnarray} \label{eq:integral_unitaries}
e^{-\beta H/2} = \int_{-\infty}^{\infty}dp R(\beta,p)e^{-iHp}
\propto \bras{0_q} e^{-iH\hat{p}} \kets{R(\beta)},
\end{eqnarray}
where $\kets{R(\beta)}=\sqrt{\beta\pi}\int_{-\infty}^{\infty}dpR(\beta,p)\kets{p}_p$.
Here we denote $\kets{p}_p$~($\kets{q}_q$) as basis of continuous variable quadrature $\hat{p}$~($\hat{q}$).
Eq.~\eqref{eq:integral_unitaries} shows a scheme that the nonunitary operator $e^{-\beta H/2}$ can be implemented on a quantum computer, assisted with an ancillary qumode. The qumode is prepared at $\kets{R(\beta)}$, evolves jointly with the system by $e^{-iH\hat{p}}$, and is finally projected onto $\kets{0}_q$. As the information related to temperature is encoded in $\kets{R(\beta)}$, we may call it as a resouce state.

\begin{figure}[h!]
	\includegraphics[width=1.0\columnwidth]{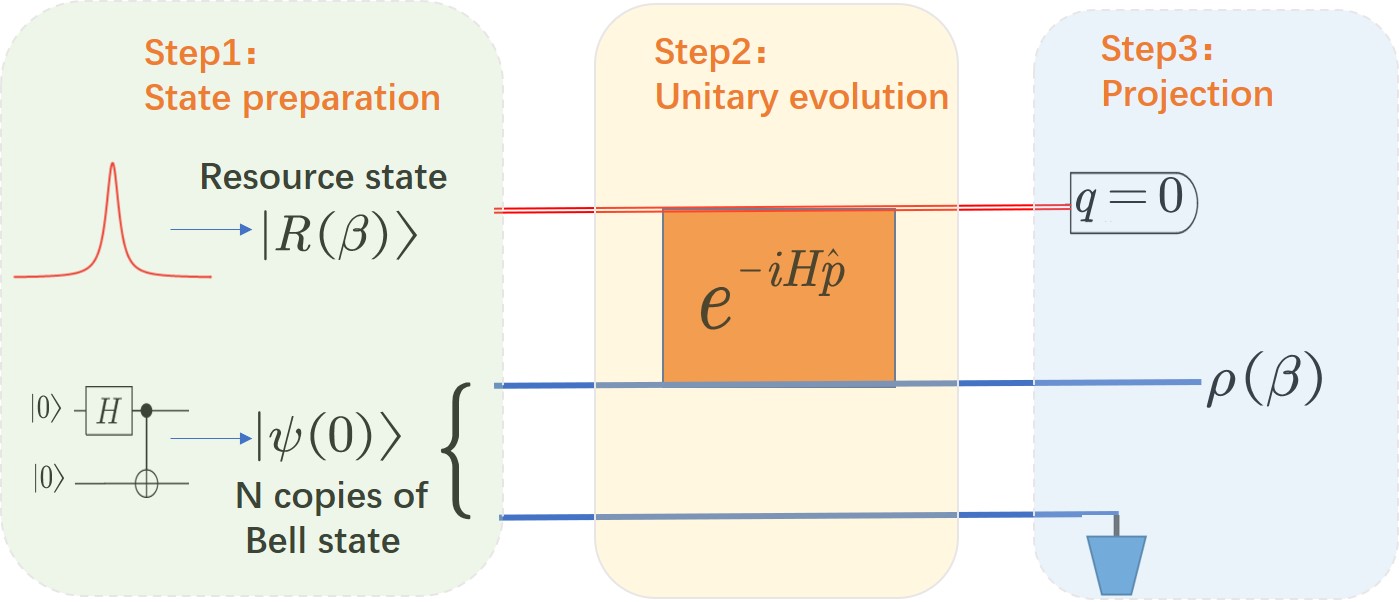}
	\caption{An illustration of preparing quantum thermal state $\rho(\beta)$ from N copies of Bell states, through coupling the system with a qumode by unitary evolution $e^{-iH\hat{p}/2}$. The qumode is initialed at the resource state $\kets{R(\beta)}$ and is finally projected onto $\kets{0}_q$.
	Thermal state $\rho(\beta)$ is obtained by discarding~(tracing out) the additional $N$ ancillary qubits.}
	\label{fig:integral_of_unitaries}
\end{figure}

\emph{Quantum algorithm.--} We now present the procedure of quantum algorithm for preparing thermal quantum simulation. The quantum algorithm is probabilistic since it postselects the quadrature $\hat{q}$ to zero. % in the homodyne detection.
In practice, the measurement should have a finite precision, which reduces the accuracy while raises the success rate. We model this effect by projecting to a squeezing state
$\kets{0,s}=s^{-\frac{1}{2}}\pi^{-\frac{1}{4}}\int dp\kets{p}_pe^{-p^2/2s^2}$,
which squeezes the quadrature $\hat{q}$ by a factor $s$. The quantum algorithm, as illustrated in Fig.~\ref{fig:integral_of_unitaries}, has three steps:
\begin{enumerate}
\item \emph{State preparation.} Prepare N copies of Bell states, $\kets{\psi(0)}=\prod_{i=1}^{N}B_{i,i+N}$, where $B_{i,i+N}=\frac{1}{\sqrt{2}}(\kets{0_i0_{i+N}}+\kets{1_i1_{i+N}})$ is a Bell state. A qumode is initialized in a resource state $\kets{R(\beta)}$. The total system is in a state $\kets{\Psi_0}=\kets{\psi(0)}\otimes\kets{R(\beta)}$. Therefore, to simulate a Hamiltonian $H$ which is encoded in $N$ qubits, our scheme needs $2N$ qubits and one qumode.	
\item \emph{Unitary evolution.} A constant is added to $H$ to make positivity of the spectrum.
 then the unitary evolution $e^{-iH\hat{p}}\otimes I$ is implemented and it couples the system with the qumode. The unitary can be decomposed with Trotter-Suzuki formula, which will be discussed later. The state turns to be
\begin{equation}
\kets{\Psi_1}=\sum_n\int_{-\infty}^{\infty}dp\phi_n(\beta,p) \kets{u_n}\otimes\kets{u^*_n}\otimes \kets{p}_p,
\end{equation}
where $\phi_n(\beta,p)=\sqrt{\frac{\beta\pi}{D}}R(\beta,p)e^{-E_np}$.
\item \emph{Projection.} Project the qumode onto a squeezing state
$\kets{0,s}$, the unnormalized final state (discarding the qumode) is
\begin{equation}
\kets{\tilde{\psi}(\beta)}=\sum_n a(E_i,\beta,s)\kets{u_n}\otimes\kets{u^*_n},
\end{equation}
where $a(E,\beta,s)=s^{-\frac{1}{2}}\pi^{-\frac{1}{4}}\int_{-\infty}^{\infty}dp\phi_n(\beta,p)e^{-\frac{p^2}{2s^2}}$.
The success rate is $O(\frac{\mathcal{Z}(\beta)}{sD})$.
\end{enumerate}

Note at $s\rightarrow\infty$ limit, $a(E,\beta,s)\propto e^{-\beta E/2}$, which exactly equals to the quantum thermal state at an inverse temperature $\beta$.

One issue of the quantum algorithm is that
the resource state can not be produced for free, e.g.,  $R(\beta\rightarrow0)$ and $R(\beta\rightarrow\infty)$ can not be efficiently prepared since they are infinitely squeezed. To solve this problem, we reveal an equivalent relation of the quantum algorithm, which says that $\rho(\beta)$ can be simulated with different pairs of resource states and unitary evolutions, namely
\begin{equation}
\left(R(\beta_0),e^{-iH\hat{p}\frac{\beta}{\beta_0}}\right) \mapsto \rho(\beta),
\end{equation}
where $\beta_0$ is adjustable. The equivalent relation is a direct consequence of an invariance under $\beta \rightarrow a\beta$ and $H \rightarrow H/a$ in Eq.~\eqref{eq:integral_unitaries}.

The equivalent relation allows us to use a fixed resource state to simulate the thermal state at varied $\beta$, which we call as adaptive TQS. From the perspective of quantum resource theory~\cite{chitambar_19}, the equivalent relation may reveal a conversion between static resource (preparing resource state) and dynamical resource~(constructing unitary $e^{-iH\hat{p}}$) and their trade-off for thermal quantum simulation.
It enriches the flexibility and feasibility of algorithmic implementation , as difficulties of resource state preparation and Hamiltonian evolution vary on different quantum platforms.

\emph{Time complexity.--}
We give a runtime analysis for a relative error $\epsilon$ in the partition function. As it is not practical to ignore the cost of preparing resource state~(e.g., at $\beta\rightarrow0,\infty$ limits), we consider adaptive TQS with a resource state $\kets{R(\beta_0)}$ that is easy to prepare. The runtime then relies on the circuit complexity of constructing the unitary operator $e^{-iH\hat{p}\beta/\beta_0}$ and the success rate.

For a Hamiltonian $H=\sum_{i=1}^Mc_i H_i$ with local terms, one can refer to Trotter-Suzuki formula~\cite{lloyd_96} to decompose $e^{-iH\hat{p}\beta/\beta_0}$.
Typically, it includes terms like evolutions of $\sigma^\alpha_i\hat{p},\sigma^\alpha_i\sigma^\gamma_{i+1}\hat{p}$ ($\alpha,\gamma=x,y,z$), which can be viewed as parity-dependent displacement operator and can be decomposed as basic gates of qubits and a hybrid gate $e^{i\theta\sigma^x\hat{p}}$~(see SM). The circuit complexity is $O(M^3\epsilon^{-1}\beta^2)$, and
may be improved with more advanced techniques ~\cite{low_17,campbell_19,childs2019theory}. For a precision $\epsilon$, the squeezing factor should be $s=O(\epsilon^{-\frac{1}{2}})$. Using amplitude amplification~\cite{Brassard_02}, the success rate becomes $O(\sqrt{\frac{\mathcal{Z}(\beta)}{sD}})$, and the algorithm should be run repeatedly with $O(\sqrt{\frac{D}{\mathcal{Z}(\beta)}}\epsilon^{-\frac{1}{4}})$ times. In total, the time complexity is $O(\sqrt{\frac{D}{\mathcal{Z}(\beta)}}M^3\beta^2\epsilon^{-\frac{5}{4}})$, which improves polynomially from methods using quantum phase estimation~\cite{poulin_09}. In addition, we found that the quantum algorithm gives a bound from below for the partition function~(see SM).

% as cite Further improvement may be made, for instance, using amplitude amplification~\cite{brassard_00}, by which repeated times can be reduced to $O(\sqrt{\frac{\mathcal{Z}(\beta)}{sd})}$. Then, the total runtime reduce to $O(L^3\sqrt{\frac{d}{\mathcal{Z}(\beta)}}\beta^{\frac{1}{2}}\epsilon^{-\frac{5}{4}})$.
%It is noted that the amplitude amplification requires operations on a continuous variable.

%
%\begin{figure}[h!]
%	\label{fig:algorithm_implement.jpg}
%	\includegraphics[width=0.8\columnwidth]{algorithm_implement.jpg}
%	\caption{Illustration of quantum algorithm and physical system to preparing quantum thermal state. (a).Quantum algorithm for preparing $\rho(\beta)$ from a completely mixed state $I/d$, by coupling the system with the qumode environment by unitary evolution $e^{-iH\hat{p}/2}$. The qumode is initialed at the resource state $\kets{R(\beta)}$ and is projected onto zero position state. (b). Illustration of a physical system with multiple qubits put in a cavity, where the cavity mode provides the continuous-variable resource state $\kets{R(\beta)}$.}
%\end{figure}

%A squeezing factor about $s=10$ is enough for our purpose, which is feasible for the recent technology.

%We focus on how the precision depends on the approximated resource state, and the squeezing factor at the projecting step. Moreover, we show that a proper choosing of specified resource state can prepare quantum thermal states at different temperature with good precision.
\begin{figure}[h!]
\includegraphics[width=1.0\columnwidth]{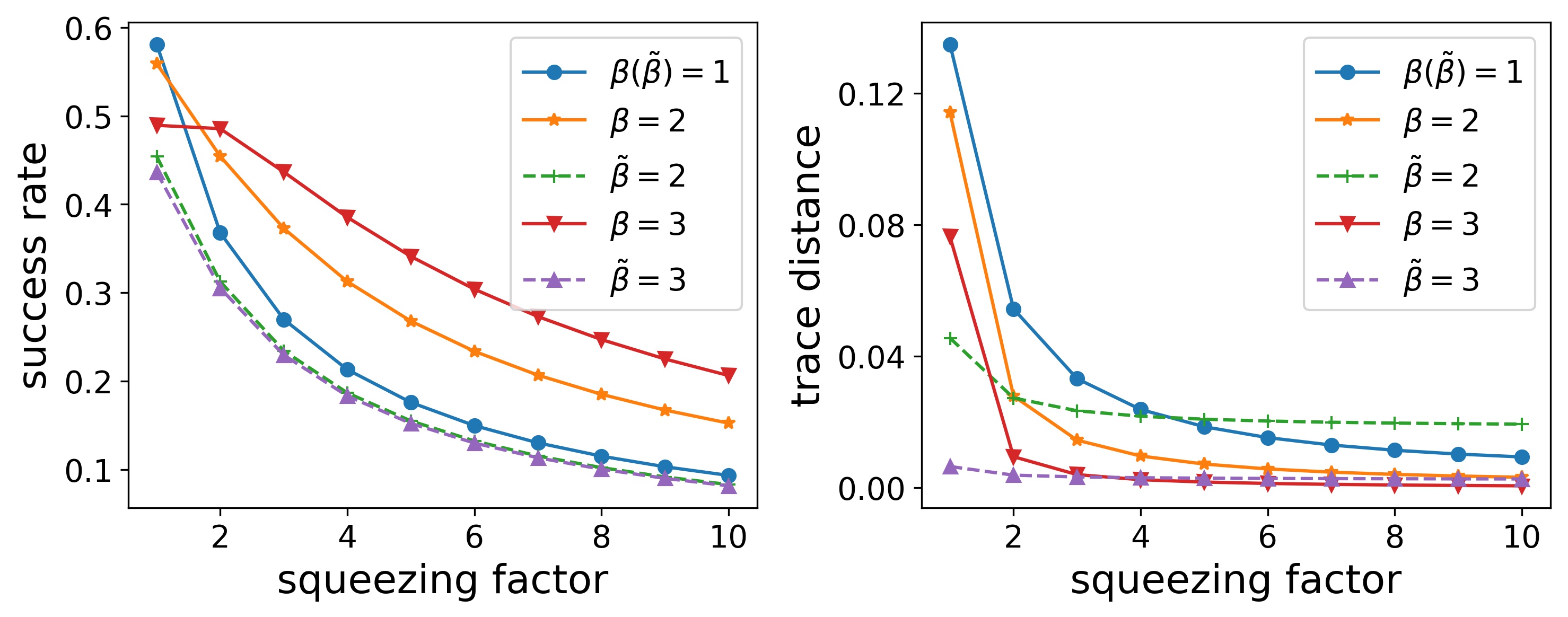}
\caption{Performance of preparing thermal states of single qubit with two different approaches: $\beta$~($\tilde{\beta}$) stands for results from TQS~(adaptive TQS).} \label{fig:single_qubit}
\end{figure}

\emph{Demonstration with single qubit.--}
To warm up, we simulate thermal state of a single qubit to highlight some features of the quantum algorithm.
For the numeral simulation, we develop a classical simulator for hybrid-variable quantum computing based on the open-source  Qutip~\cite{johansson_13}.
The Hamiltonian is $H=g(\sigma_x+c)$~(here $g>0$ and $c=1$ enforces all eigenvalues are nonnegative). After initialing the cavity mode into a resource state and preparing a Bell state, an unitary $e^{-ig(\sigma^x+1)\hat{p}t}$ performs on the cavity mode and the system qubit, and finally the cavity mode is squeezed and projected onto zero photon. We use two approaches for preparing thermal states at $\beta=1,2,3$. The first fixes $t=1$ and uses resource states $\kets{R(\beta)}$ with $\beta=1,2,3$, respectively. The second is adaptive TQS, fixing the resource state $\kets{R(\beta=1)}$ and using $t=1,2,3$, respectively. Due to finite squeezing at the projection, those simulated thermal states can only approximate the exact thermal states. We use trace distance to measure the precision,  $d_r(\beta)=\frac{1}{2}\text{Tr}|\tilde{\rho}(\beta)-\rho(\beta)|$.
In Fig.~\ref{fig:single_qubit},
we can see that trace distances decrease rapidly with an increase of squeezing factor, at the cost of decreasing success rate, which are expected. Using adaptive
TQS, the increasing of precision can be faster, and the trace distance can be small even at small $s$. Moreover, a moderate truncation of phonon number~(e.g., up to $7$ even-number Fock states) can reach a good precision~(see SM). The above results indicate that we can use an adaptive approach with approximated resource state for thermal quantum simulation.

\begin{figure}[h!]
	\includegraphics[width=1.0\columnwidth]{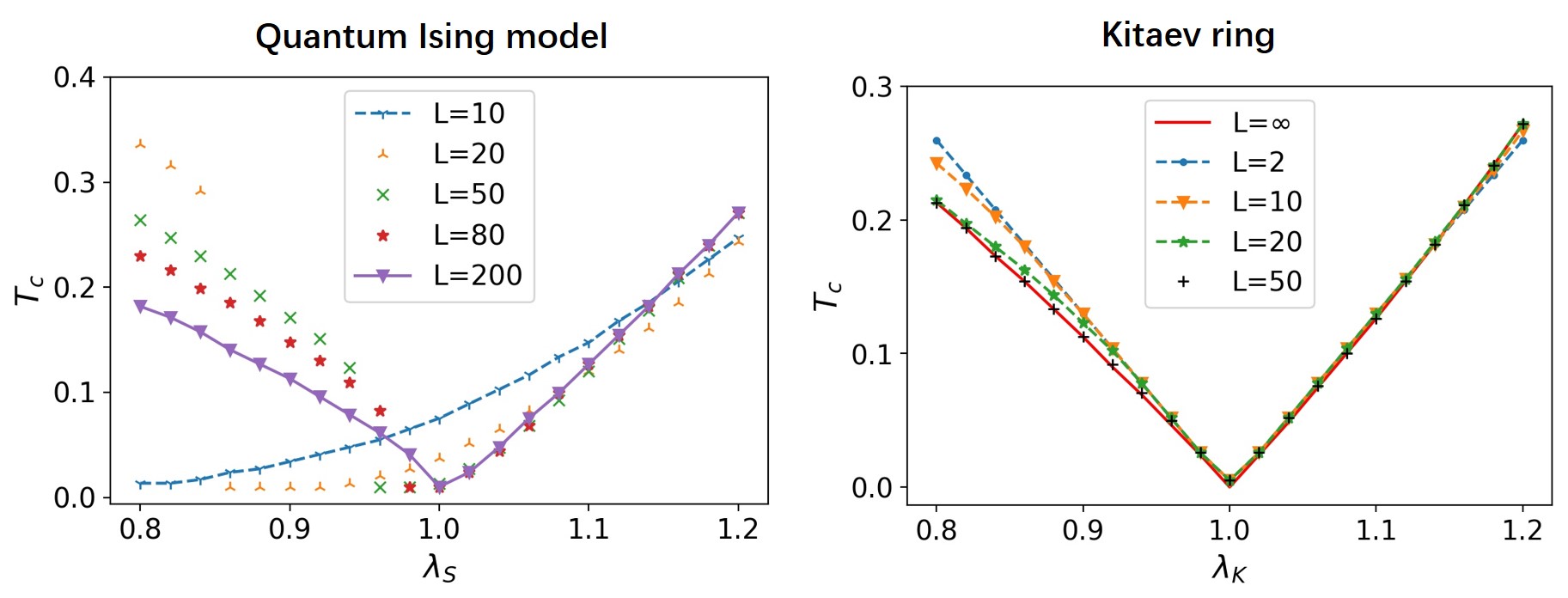}
	\caption{Finite temperature phase diagram of the one-dimensional Kitaev ring and quantum Ising model.}
	\label{fig:ising_kitaev}
\end{figure}

\emph{Simulation of the finite-temperature phase diagram.--}
%Now we are ready to simulate some important physics that thermal fluctuation plays a significant role. It is expected that a distribution on low-energy eigenstates can not be ignored when thermal fluctuation excesses the energy gap. For a quantum phase transition~(QPT) point that is gapless, a quantum critical regime emerges as a profound consequence of an interplay between quantum and thermal fluctuations~\cite{Sachdev}. The quantum critical regime can be characterized by scaling behavior of physical quantities and an universal crossover temperature, which often forms a $V$-shape curve around the QPT point. It is challenge, however, to simulate a scaling behavior within the current quantum technique, for a lack of well controlling of a large number of qubits at different temperatures. On the other hand, crossover temperature corresponds to a temperature maximizing correlations of some physical quantities, such as magnetic susceptibility, which can be easier to simulate. Thus, our goal is to employ the adaptive TQS for quantum criticality and show it can reveal an universal crossover temperatures.
%%%%%%%%%%%%%%%%%%%%%%%%%%%%%%%%%%%
We now consider our TQS with two textbook models, the Kitaev ring (a spinless p-wave superconductor) ~\cite{kitaev2001unpaired} and quantum Ising model ~\cite{Sachdev}. The Hamiltonian of the Kitaev ring reads
\begin{equation}\label{ham:kitaev}
H_{K}=-J\sum_{i=1}^{L}(c_i^\dagger c_{i+1}+c_i^\dagger c_{i+1}^\dagger + h.c.)-\mu\sum_{i=1}^{L} c_i^\dagger c_i,
\end{equation}
where fermionic operators $c_{L+1}=c_{1}$ as we consider the periodic condition.
The model has a topological phase transition at $\lambda_K\equiv\mu/2J=1$. Using Jordan-Wigner transformation, $
c_i=\prod_{j=1}^{i-1}\sigma_j^z\sigma_i^-,~
c_i^\dagger=\sigma_i^+\prod_{j=1}^{i-1}\sigma_j^z,~c_i^\dagger c_i=\frac{1}{2}(\sigma_i^z-1)$,
the Hamiltonian of the Kitaev ring can be mapped into a spin model
 \begin{equation}\label{eq:kitaev_JW}
H_{S}=-h\sum_{i}^{L}\sigma_i^z-J\sum_{i}^{L-1}\sigma_i^x\sigma_{i+1}^x-J\sigma_1^yP_L\sigma_L^y+E_0,
\end{equation}
 where $h=-\frac{\mu}{2}$, $P_L=\prod_{i=2}^{L-1}\sigma_i^z$ is a string operator, $E_0$ is added for assuring nonnegative spectrum~(which is demanded for the quantum algorithm).
 %%%%%%%%%%%%%%%%%%%%%%%%%%%%%%%%%%%%%%%%%%%%%%%%%%
%  and $H_S$ is the Hamiltonian of the quantum Ising model given by
%\begin{equation}\label{eq:kitaev_JW}
%H_{S}=-h\sum_{i}^{L}\sigma_i^z-J\sum_{i}^{L-1}\sigma_i^x\sigma_{i+1}^x
%\end{equation}
% with $h=-\frac{\mu}{2}$. The model has a phase transition at $\lambda_S\equiv h/J=1$ .
The spin model $H_S$ become the quantum Ising model if $\sigma_1^yP_L\sigma_L^y$ is replaced by $\sigma_L^x\sigma_1^x$, and it has a phase transition at $\lambda_S\equiv h/J=1$. This difference of the Ising and Kitaev models has a big effect on the quantum criticality at small sizes, as addressed below.
The finite-temperature phase diagrams of these two models in the infinite lattice size are the same, which have an important V-shape crossover structure \cite{Sachdev} as shown in Fig. \ref{fig:ising_kitaev}.

From numeral calculation~(see SM for details), it is shown that almost $L\approx80$ qubits is required for the quantum Ising chain to show well shaped crossover temperatures, as seen in Fig~\ref{fig:ising_kitaev}, and $L$ should be larger closer to the QPT point. For the Kitaev ring, in contrast, the temperature crossovers are very close in shape for a quite large range even for small $L$. The subtle lies in that Kitaev ring always has an energy gap $\Delta=J|1-\lambda_K|$, while the quantum Ising model will have low-lying in-gap states when $\lambda_S<1$, leading to small crossover temperatures for $\lambda_S<1$, which is obvious for $L=10,20$. Thus, we refer to a small size Kitaev ring for simulating finite-temperature phase diagram with quantum computers.
%At $L=2$, $H_{\text{spin}}$ reduces to $
%H_{2s}=-J(\sigma_1^x\sigma_1^x+\sigma_1^y\sigma_1^y)-h(\sigma_1^z+\sigma_2^z)$.

%(note that $\chi_m$ can be obtained also from measuring two-spin correlations of $\rho(\beta)$).

\begin{figure}[h!]
	\includegraphics[width=1.0\columnwidth]{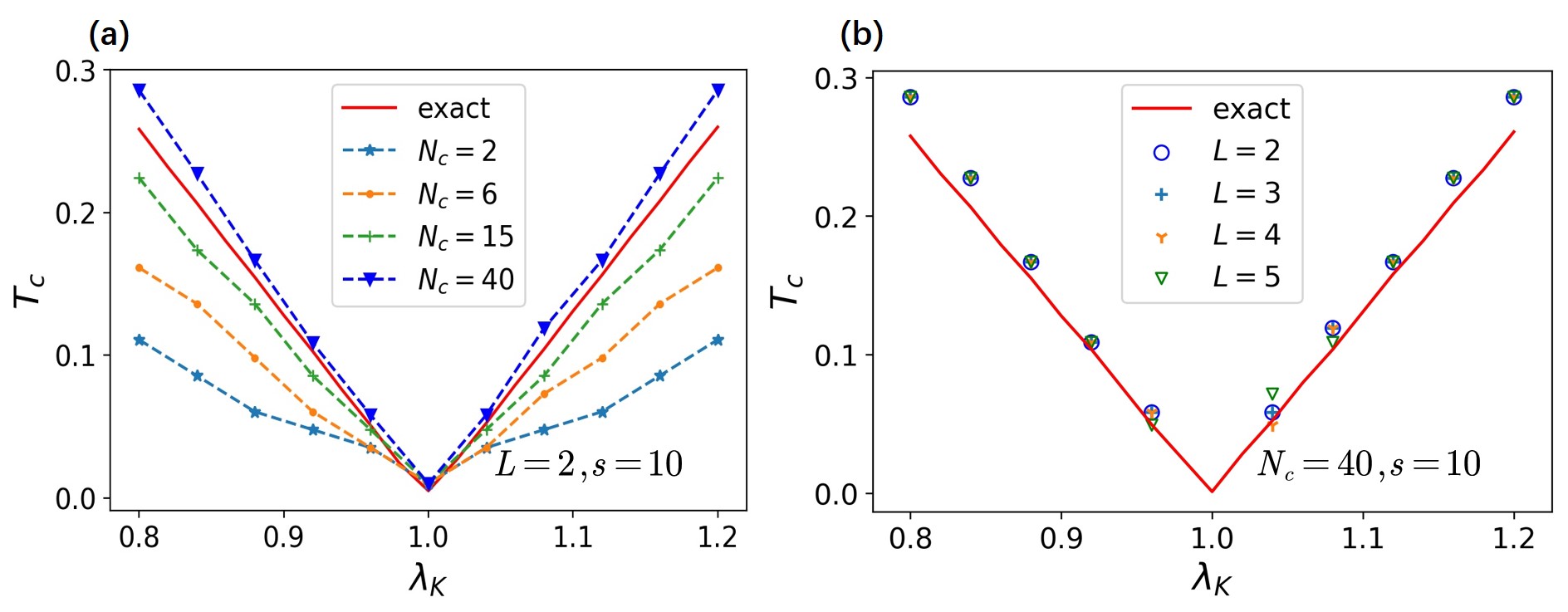}
	\caption{Simulation of crossover temperature for the Kitaev ring in the quantum critical regime with adaptive TQS. (a). Crossover temperatures that use different truncating of photon number $N_c$ for the resource state $\kets{R(\beta=4)}$.(b).Crossover temperature for different lattice size, $L=2,3,4,5$.}
	\label{fig:T_crossover}
\end{figure}

To address the feasibility of the physical implementation, we adopt adaptive TQS for simulating thermal states of the Kitaev ring, using a resource state $\kets{R(\beta=4)}$ and a squeezing factor $s=10$.  The  crossover temperature for each $\lambda_K$ is determined by the temperature that maximizes the magnetic susceptibility~$\chi_m$~(corresponding to fluctuation of fermion number in the Kitaev ring), which can be obtained by measuring the magnetization $\mathcal{M}=\frac{1}{L}\sum_i\text{Tr}\rho(\beta)\sigma_i^z$ and then calculating the magnetic susceptibility $\chi_m=\frac{\partial \mathcal{M}}{\partial h}$ using a finite difference. The result is shown in Fig.~\ref{fig:T_crossover}.
For $L=2$, we compare results of different truncation for the resource state~($N_c$ is the number of Fock basis of even-number photons), and it can be seen that the $V$-shape crossover temperature approaches the exact one when increasing $N_c$. We also simulate $L=2,3,4,5$, which demonstrates very close crossover temperatures. Therefore, a remarkable advantage of our quantum algorithm is that observation of the important crossover temperature phase diagram of the Kitaev ring needs only a few qubits.

\emph{Experimental realization.--}
We now discuss physical implementation of the quantum algorithm, which relies on hybrid variable quantum computing. Promising candidates include %optical system~\cite{furusawa2011quantum},
trapped ions~\cite{leibfried_03,zhu_06,häffner_08,monroe_13,zhang_18} and superconducting circuits~\cite{krastanov_15,li_17,hofheinz_08,heeres_15,li_17,hu_19,song_19}, etc. We take superconducting circuit system as an example, for their well controllable CV cavity mode and its coupling to the qubits. The scheme can be straightforwardly used in trapped ion quantum computer.
%We outline basic components of implementing with superconducting, following the main procedures of the quantum algorithm.

Firstly, the resource state can be expanded in a Fock space,
$\kets{R(\beta)} =\sum_{n=0}^{\infty}r_{n}\kets{2n}$, where
$r_{n}=\int_{-\infty}^{\infty}dq\sqrt{\frac{\beta}{2}}e^{-\beta|q|/2} H_{2n}(q)e^{-\frac{q^2}{2}}$ and $H_{n}(q)$ is the $n$-th order Hermite function.
With a truncation of photon number, the state can be prepared in a cavity by superposing even-number photons.
This can be achieved with a sequence of qubit rotation and Jaynes-cumming type qubit-cavity coupling ~\cite{law_96,hofheinz_09}, or using the number-dependent arbitrary phase gate and displacement operator in the dispersive regime~\cite{heeres_15,wang_17}. Note that resource states around $\beta=2$ can be approximated with a few components of small number photons, and are feasible in both approaches. Thus, we can chose one such resource state and use it for adaptive TQS.
Secondly, construction of $e^{-iH\hat{p}t}$ can be compiled into basic qubit gates and only one hybrid variable gate $e^{i\theta\sigma^x\hat{p}}$, as discussed before. All are standard quantum operations in superconducting circuit system, and remarkably, the hybrid gate $e^{i\theta\sigma^x\hat{p}}$ can be readily realized in the strong coupling limit ~\cite{niemczyk_10,forn-díaz_17,forn-diaz_19}. Thirdly,
projection onto a squeezing state $\kets{0,s}$ can be implemented by first performing a squeezing on the CV mode, and then post-selecting the CV mode to the vacuum state~(zero photon state). Further, we can measure the system to access the
thermal state by quantum state tomography, or studying quantum statistical mechanism by measuring physical quantities, such as heat capacity, magnetic susceptibility, etc. As for the Kitaev ring, we note that evolution of nonlocal term $\sigma_1^yP_L\sigma_L^y\hat{p}$ can be constructed efficiently~(see SM). Thus, all above well-developed quantum operations can render a feasible implementing protocol for thermal quantum simulation of the Kitaev ring, in order to illuminate the novel quantum critical regime on small quantum processors.

%While we exemplify the implementation with superconducting circuit, we remark especially that trapped ions may also be used for implementing the quantum algorithm for thermal quantum simulation, for its well-controllable motional modes.

%\emph{Discussion and summary.--}
\emph{Summary.--}
%Our work uncover a paradigm for simulating finite temperature quantum systems, which incorporates quantum resources of both auxiliary continuous variables and dynamics of the system. With an integral-of-unitaries, the quantum algorithm can be constructed rather straightforwardly. We expect such a paradigm may be extended for quantum simulation of physical systems out-of thermal equilibrium.
We have proposed a quantum algorithm for thermal quantum simulation assisted with an auxiliary CV resource state.
We have confirmed its power by simulating finite-temperature phase diagram of the Kitaev ring, and found that the important crossover phase diagram of the
model can be accurately determined by a quantum computer with only
a few qubits. Thus, our work may pave the way for studying finite temperature quantum systems in experiments.

\begin{acknowledgments}
	This work was supported by the Key-Area Research and Development Program
	of GuangDong Province (Grant No. 2019B030330001), the National Key Research
	and Development Program of China (Grant No. 2016YFA0301800), the National Natural Science Foundation of China (Grants No. 91636218 and
	No. U1801661), the Key Project of Science and Technology of Guangzhou (Grant No. 201804020055), and the CRF of Hong Kong (No. C6005-17G).
\end{acknowledgments}

%\bibliography{gibbs}

\onecolumngrid
\appendix{
\section{Supplementary Material}	
\section{Thermofield double state.}	
We prove that the purified thermal state at finite temperatures can be written as $N$ pairs of Bell states.
Denote $\{E_n\}$ and $\{\kets{u_n}\}$~($n=0,1,...,2^N-1$) as eigenvalues and eigenstates for a Hamiltonian $H$ which can be encoded with $N$ qubits. The system of $H$ at infinite temperatures is described by
a complete mixed state. To purify this complete mixed state, we use a thermofield double~(TFD) state at infinite temperatures,
\begin{equation}\label{eq:thermofield}
\kets{\psi(\beta=0)}\equiv\frac{1}{\sqrt{D}}\sum_n\kets{u_n}\otimes\kets{u^*_n}=\frac{1}{\sqrt{D}}\sum_n\kets{n}\otimes\kets{n},
\end{equation}
where $D=2^N$ is the dimension of Hilbert space.
We now prove the equation.
Note that both $\{\kets{u_n}\}$ and $\{\kets{n}\}$ form a complete set of basis for the Hilbert space of $N$ qubits, we can find an unitary transformation that $\kets{u_n}=\sum_n U_{nm}\kets{m}$.
Then,
\begin{eqnarray}
\sum_n\kets{u_n}\otimes\kets{u^*_n}&=&\sum_{nmm'} U_{nm}U^*_{nm'}\kets{m}\otimes\kets{m'} \nonumber \\
&=&\sum_{mm'}\delta_{mm'}\kets{m}\otimes\kets{m'}\nonumber \\
&=&\sum_{n}\kets{n}\otimes\kets{n}
\end{eqnarray}

We prove that  $\kets{\psi(0)}=\prod_{i=1}^{N}B_{i,i+N}$, where $B_{i,i+N}=\frac{1}{\sqrt{2}}(\kets{0_i0_{i+N}}+\kets{1_i1_{i+N}})$.
With a binary expression $n=n_1n_2..,n_{N}$~($n_i=0,1$), we can write,
\begin{eqnarray}
\kets{\psi(\beta=0)}&=&\frac{1}{\sqrt{D}}\sum_{n}\kets{n}\otimes\kets{n} \nonumber \\
&=&\frac{1}{\sqrt{D}}\sum_{n_1n_2...n_N}\kets{n_1n_2...n_N}\otimes\kets{n_1n_2...n_N}\nonumber \\
&=&\frac{1}{\sqrt{D}}\prod_{i=1}^{N}\sum_{n_i=0,1}\kets{n_i}_i\otimes\kets{n_i}_{i+N}\nonumber \\
&=&\prod_{i=1}^{N}\frac{1}{\sqrt{2}}(\kets{0_i0_{i+N}}+\kets{1_i1_{i+N}})
\end{eqnarray}

It should be reminded that other purification protocols are possible, for instance, $\kets{\psi(\beta=0)}\equiv\frac{1}{\sqrt{D}}\sum_n\kets{u_n}\otimes\kets{n}$. 
%For finite temperature, as the entanglement is not maximal, it is expected that an auxiliary system can have less qubits than the system studied. 
%How to optimize the purification has gained some attention recently~\cite{nguyen_18}.

\section{Exact free energy}
For the Kitaev ring and the quantum Ising model under periodic boundary condition~(PBC), we give methods to exactly calculate their free energy at different parameters and temperatures. Specified attention should be given to the quantum Ising model, since its energy spectrum should be calculated in the even-parity and odd-parity subspaces, respectively. From the free energy the magnetic susceptibility can be obtained.

\textbf{Kitaev ring}.
For the Kitaev ring, the Hamiltonian after diagonalization in k-space with Bogoliubov transformation can be written as,
\begin{equation}
H_K=\sum_{i=0}^{L-1}\xi(k_i)(\eta_{k_i}^\dagger\eta_{k_i}-\frac{1}{2}),
\end{equation}
where the single-particle spectrum is $\xi(k)=2J\sqrt{1+\lambda_K^2-2\lambda_K\cos{k}}$, and we set $\lambda_K=2\mu/J$.
For this non-interacting fermionic model that follows Fermi-Dirac statistics, the partition function can be easily calculated as,
\begin{equation}\label{eq:Z_kitaev}
Z(\beta)=\prod_{i=0}^{L-1}(e^{-\beta\xi(k_i)/2}+e^{\beta\xi(k_i)/2}).
\end{equation}
Then, the free energy per site is,
\begin{equation}
f(\beta)=-\beta^{-1}\log{Z(\beta)}/L=-\beta^{-1}\left[-\frac{1}{L}\sum_{i=0}^{L-1}\log{\cosh{\frac{\beta\xi(k_i)}{2}}}-\log{2}\right].
\end{equation}
At the thermodynamical limit, it turns to be,
\begin{equation}
f(\beta)=-\beta^{-1}\log{Z(\beta)}/L=-\beta^{-1}\left[\frac{1}{2\pi}\int_{0}^{2\pi}dk\log{\cosh{\frac{\beta\xi(k)}{2}}}-\log{2}\right].
\end{equation}

\textbf{Quantum Ising model}. The quantum Ising model under PBC is,
\begin{equation}\label{eq:kitaev_JW}
H_{I}=-h\sum_{i}^{L}\sigma_i^z-J\sum_{i}^{L-1}\sigma_i^x\sigma_{i+1}^x,
\end{equation}
where $\sigma_{L+1}^\alpha=\sigma_{1}^\alpha$.
Using the Jordan-Wigner transformation,
it becomes a fermionic Hamiltonian,
\begin{equation}\label{ham:ising_jw}
H_I=-J\sum_{i=1}^{L-1}(c_i^\dagger c_{i+1}+c_i^\dagger c_{i+1}^\dagger + h.c.)-\frac{h}{2}\sum_{i=1}^{L} c_i^\dagger c_i-J(-1)^{\hat{N}}(c_L-c_L^\dagger)(c_1+c_1^\dagger),
\end{equation}
where $\hat{N}=\sum_{i=1}^{L}c_i^\dagger c_i$. Compared with the Kitaev ring, there is a global phase $(-1)^{\hat{N}}$ that depends on the parity of fermion number. Thus, we should discuss even and odd parity subspace separately.
\begin{enumerate}
	\item Even-parity subspace. We should have $c_{L+1}=-c_1$ in accordance with the third term of Eq.~\eqref{ham:ising_jw}. This enforces $k_i$ to take values $k_i=\frac{2\pi i}{L}+\pi/L$,($i=0,1,...,L-1$). By Bogoliubov transformation, the Hamiltonian can be diagonalized as,
	\begin{equation}
	H^e_I=\sum_{i=0}^{L-1}\xi(k_i)(\eta_{k_i}^\dagger\eta_{k_i}-\frac{1}{2}),
	\end{equation}
	where $\xi(k)=2J\sqrt{1+\lambda_S^2-2\lambda_S\cos{k}}$, and we have introduced $\lambda_S=h/J$. It should be emphasized that only even number of excitations can allow for this Hamiltonian due to the constraint of even-parity subspace. This makes that a formula as in Eq.~\eqref{eq:Z_kitaev} can not apply here.
	
	\item Odd-parity subspace. We have $c_{L+1}=c_1$, and $k_i$ to take values $k_i=\frac{2\pi i}{L}$,($i=0,1,...,L-1$).
	The Hamiltonian in this subspace writes,
	\begin{equation}
	H^o_I=\sum_{i=1}^{L-1}\xi(k_i)(\eta_{k_i}^\dagger\eta_{k_i}-\frac{1}{2})-2J(1-\lambda_S)(c_0^\dagger c_0-\frac{1}{2}).
	\end{equation}
	The last term comes from the fact that a spinless p-wave pairing can not happen at $k=0$. Similarly, only odd-number excitations are allowed for the Hamiltonian $H^o_I$.
\end{enumerate}

With Hamiltonians in both even-parity and odd-parity subspaces, we can get all eigenstates for the quantum Ising model, and then calculate the partition function. However, this can be intractable when $L$ is large, e.g., $L=50$, because there is an exponential growth number of terms to be added.

We use a recursion relation for the partition function which is scalable for large $L$.
Let $J$ and $G$ are contributions to the  partition function from the even and odd parity subspace, separately. Then, $Z=J+G$. We  discuss how to calculate $J$, and $G$ can be calculated similarly. Denote $J^e_n$~($J^o_n$) as a partition function that comes from even~(odd) number single-particle excitations for modes in $\eta_{k_i}^\dagger$ that $0\leq i\leq n$. Then, we have a recursion relation,
\begin{equation}
J^e_n=J^e_{n-1}+e^{-\beta\xi(k_n)}J^o_{n-1},~~~J^o_n=e^{-\beta\xi(k_n)}J^e_{n-1}+J^o_{n-1},~~~
J^e_1=e^{-\beta E_0}, ~~~J^o_1=e^{-\beta E_0}e^{-\beta\xi(k_0)},
\end{equation}
where $E_0=-\frac{1}{2}\sum_{i=0}^{L-1}\xi(k_i)$ is the ground state energy. By writing the recursion relation in a matrix form, we can  calculate $J=J^e_L$ very efficiently. Note that for the odd-parity subspace we should get $G=G^o_L$, correspondingly.

With obtained partition function $Z=J+G$, the free energy can be got using $F=-T\log{Z}$. In practice, we can take $e^{-\beta E_0}$ outside the expression $Z$ since it can be a very large number.

\section{More simulation results.}
\textbf{Effect of truncation of photon number}.For practice, we consider a truncation of the Fock space and the resource state thus superposes a finite set of photon numbers. We show that a moderate truncation is enough for thermal quantum simulation with good accuracy, which is shown in Fig.~\ref{fig:single_qubit_cut}, illustrated by thermal state of a single qubit.
\begin{figure}[H]
	\includegraphics[width=1.0\columnwidth]{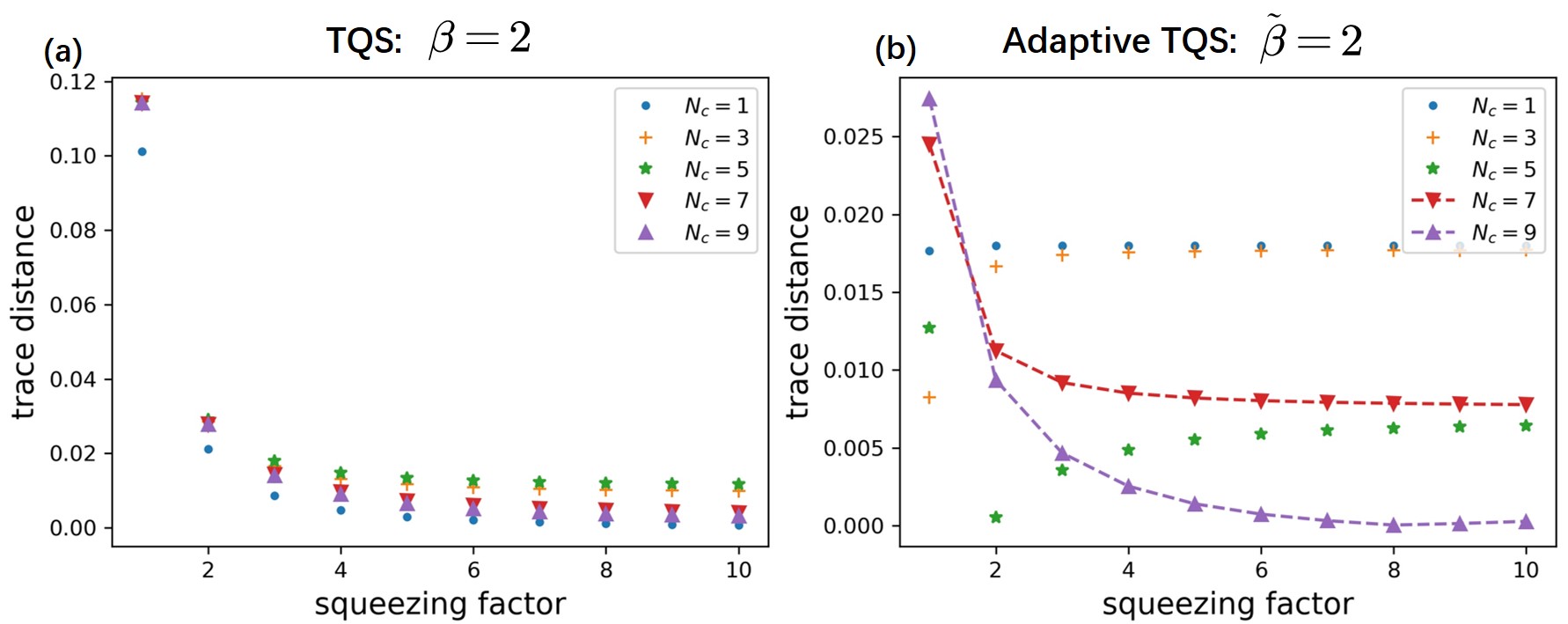}
	\caption{Effects on truncation of resource state in the Fock space for preparing thermal states of single qubit with two different approaches: $\beta$~($\tilde{\beta}$) stands for results from TQS~(adaptive TQS).
		Here, $N_c$ is the number of Fock basis of even-number photons, for instance, for $N_c=4$, $\kets{0},\kets{2},\kets{4},\kets{6}$ are used.
	} \label{fig:single_qubit_cut}
\end{figure}

\textbf{Free energy}. Accuracy simulation of free energy is key for thermal numeral simulation. Here, we present numeral results for free energies for the Kitaev ring for different $\lambda$ and $T$, as well as how it depends on the squeezing factor.
As seen in Fig.~\ref{fig:free_energy}, free energy can be simulated very accurately. Moreover, free energies at small squeezing factors are larger than exact results, which is consisted to theoretical analysis that says the quantum algorithm gives a bound from above for the free energy.
\begin{figure}[H]
	\includegraphics[width=1.0\columnwidth]{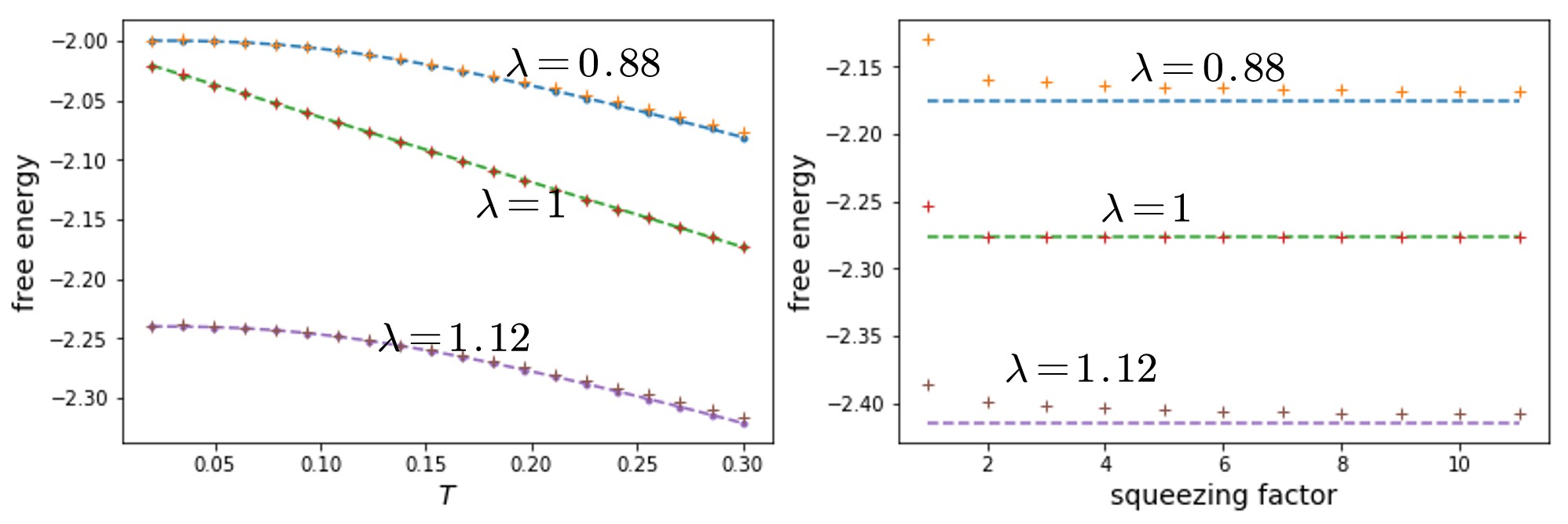}
	\caption{Free energies for the Kitaev ring at $L=2$. Dash lines correspond to exact results and markers are results from adaptive TQS. The left figure is obtained for a squeezing factor $s=10$, and the right figure is obtained at temperature $T=0.4$. All are calculated under a truncation of $N_c=35$.
	} \label{fig:free_energy}
\end{figure}

\section{Decomposition of hybrid-variable quantum operators.}
To simulate $e^{-iH\hat{p}t}$, we use a trotterization to decompose it into
short time evolutions, evolving operators $e^{-i\delta \sigma^i_z\hat{p}}$,
$e^{-i\delta \sigma^i_x \sigma^{i+1}_x\hat{p}}$, $e^{-i\delta \sigma^{1}_yP_L\sigma^{L}_y\hat{p}}$, and so on. We show that all those operators can be compiled into basic quantum gates of qubits and only one qubit-qumode hybrid quantum gate $e^{-i\delta \sigma_x \hat{p}}$, as illustrated in Fig.~\ref{fig:gate_decom}, which is similar to the technique in the textbook~\cite{nielsen_chuang_2010}. For convenience, we also denote $X=\sigma^x, Y=\sigma^y, Z=\sigma^z$. Here, we can use some basic single-qubit transformations,
\begin{equation}
HZH=X,~~~ HXH=Z,~~~ HYH=-Y,~~~SXS^\dagger=Y.
\end{equation}
And those evolve CNOT gates,
\begin{equation}
CZ_2C=Z_1Z_2,~~~CZ_1C=Z_1,~~~,CX_1C=X_1X_2, ~~~CX_2C=X_2,
\end{equation}
where we denote $C$ as CNOT gate with qubit 1 as control and qubit 2 as target.
We remark $e^{-i\delta \sigma^i_x \sigma^{i+1}_x\hat{p}}$ can be seen as parity-dependence displacement operator, where the direction of displacement depends on the parity $\sigma^i_x \sigma^{i+1}_x$.

For the Kitaev ring, specifically evolutions of $\sigma_1^yP_L\sigma_L^y\hat{p}$ can be decomposed, following the method in Fig.~\eqref{fig:gate_decom}, where we have shown the case $L=2,3$. For other $L$, one can use an identity,
\begin{equation}
e^{-i\delta\sigma_1^y\sigma_2^z...\sigma_L^y\hat{p}}=S_1H_1S_LH_LHe^{-i\delta\sigma_1^z\sigma_2^z...\sigma_L^z\hat{p}}H_1H_LS_1^\dagger S_L^\dagger,
\end{equation}
where $e^{-i\delta\sigma_1^z\sigma_2^z...\sigma_L^z\hat{p}}$ can be constructed with the same fashion as $e^{-i\delta\sigma_1^z\sigma_2^z}$ and $e^{-i\delta\sigma_1^z\sigma_2^z\sigma_3^z}$.

\begin{figure}[H]
	\includegraphics[width=0.8\columnwidth]{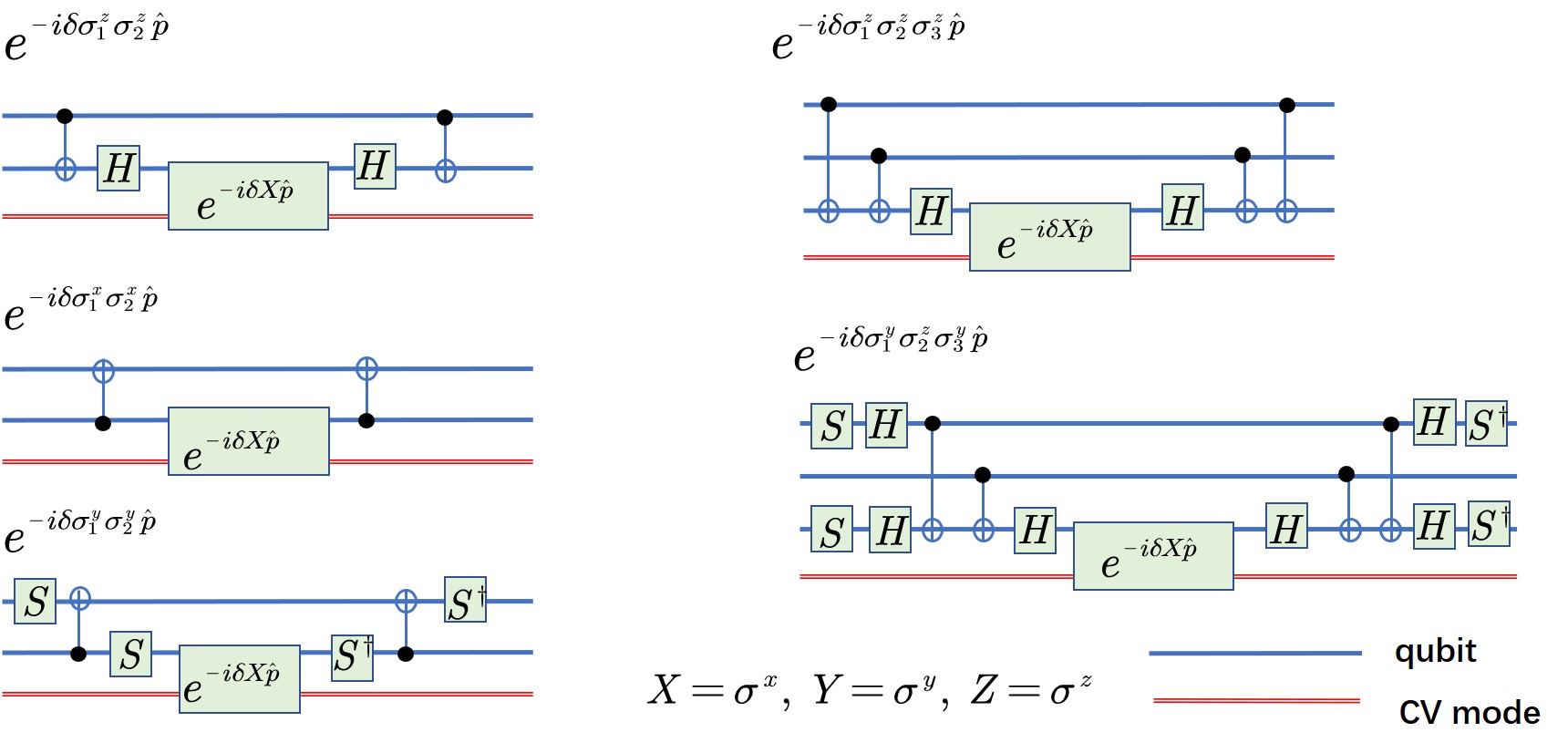}
	\caption{Decomposition of hybrid quantum gates with a series of CNOT and a qubit-qumode gate $e^{-i\delta \sigma_x \hat{p}}$.}
	\label{fig:gate_decom}
\end{figure}

\section{Requirement of squeezing factor}\label{app:error_analysis}
The task is to calculate the requirement of $s$ with the required accuracy $\epsilon$ for simulating quantum thermal state with the inverse temperature $\beta$. For this we use a relative accuracy $\epsilon$ of both  the partition function and the quantum Gibbs distribution. It is expected that the former is less demanding for squeezing factor under relative accuracy  $\epsilon$.  In the following, we find that both may give the same dependence of  the squeezing factor $s$ on $\epsilon$. We first consider TQS for $\beta$ using $\kets{R(\beta)}$, assuming that an arbitrary resource state can be prepared. Then, we consider adaptive TQS with an easy-to-prepare resource state.

A relative accuracy $\epsilon$ of the partition function is defined as
\begin{equation}
e(\beta,s):=\frac{|\tilde{\mathcal{Z}}(\beta,s)-\mathcal{Z}(\beta)|}{\mathcal{Z}(\beta)}\leq \epsilon.
\end{equation}
Here $\tilde{\mathcal{Z}}(\beta,s) \propto \int_{E_\text{0}}^{E_\text{1}}a^2(E,\beta,s)D(E)dE$, where $D(E)$ is spectral density of the system. Note that $\tilde{\mathcal{Z}}(\beta,s)-\mathcal{Z}(\beta)$ can be approximated as
\begin{eqnarray} \label{eq:app_numerator_Z}
&&\int_{E_\text{1}}^{E_\text{0}}D(E)dE\int_{-\infty}^{\infty} dp \frac{2}{\pi}\frac{\beta}{\beta^2+4p^2}(e^{-\frac{p^2}{2s^2}}-1)e^{-iEp}\nonumber \\
&&\times\int_{-\infty}^{\infty} dp \frac{2}{\pi}\frac{\beta}{\beta^2+4p^2}(e^{-\frac{p^2}{2s^2}}+1)e^{-iEp}\nonumber\\
&&=-16\int_{E_\text{0}}^{E_\text{1}}dED(E)(e^{\frac{\beta^2}{8s^2}}-1)(e^{\frac{\beta^2}{8s^2}}+1)e^{-\beta E},
\end{eqnarray}
where in the third line we have applied Cauchy's residue theorem in complex analysis, with an integral of $p$ in the half complex plane enclosing the pole $p=-i\beta/2$ (the other pole is not used since the integral must depend on $E$). We note that Eq.~\eqref{eq:app_numerator_Z} is small than zero, which suggests that the quantum algorithm would give a bound from below for the partition function, and consequently a bound from above for the free energy.

For small $s/\beta$, using $e^{x}\approx 1+x$, then
\begin{eqnarray} \label{eq:error_simpled}
e(\beta,s) \sim \frac{2\beta^2}{s^2}\frac{\int_{E_\text{0}}^{E_\text{1}}dED(E)e^{-\beta E}}{\int_{E_\text{0}}^{E_\text{1}}dED(E)e^{-\beta E}} = \frac{2\beta^2}{s^2}
\end{eqnarray}
It is interesting to note that the final expression is independent of the details of the spectral function. For TQS of $\beta$ using a resource state $\kets{R(\beta)}$,
to achieve $e(\beta,s)\leq \epsilon$ at inverse temperature $\beta$, the required squeezing factor is $s \sim \beta \epsilon^{-\frac{1}{2}}$.

We now consider adaptive TQS at inverse temperature $\beta$ with a resource state $\kets{R(\beta_0)}$. Eq.~\eqref{eq:app_numerator_Z} then is,
\begin{eqnarray} \label{eq:app_numerator_Za}
&&\int_{E_\text{1}}^{E_\text{0}}D(E)dE\int_{-\infty}^{\infty} dp \frac{2}{\pi}\frac{\beta_0}{\beta_0^2+4p^2}(e^{-\frac{p^2}{2s^2}}-1)e^{-iEp\frac{\beta}{\beta_0}} \nonumber \\
&&\times\int_{-\infty}^{\infty} dp \frac{2}{\pi}\frac{\beta_0}{\beta_0^2+4p^2}(e^{-\frac{p^2}{2s^2}}+1)e^{-iEp\frac{\beta}{\beta_0}}\nonumber\\
&&=-16\int_{E_\text{0}}^{E_\text{1}}dED(E)(e^{\frac{\beta_0^2}{8s^2}}-1)(e^{\frac{\beta_0^2}{8s^2}}+1)e^{-\beta E}.
\end{eqnarray}
One can get $s \sim \epsilon^{-\frac{1}{2}}$, which is independent of $\beta$. This can explain why adaptive TQS is less demanding for squeezing. The cost, however, is that the complexity to construct the unitary evolution increases with a factor $\beta^2$.

For a relative error $\epsilon$ of the Gibbs distribution, it requires that
\[
e(\beta,s):=\frac{|a^2(E,\beta,s)/\tilde{\mathcal{Z}}(\beta,s)-e^{-\beta E}/\mathcal{Z}(\beta,s)|}{e^{-\beta E}/\mathcal{Z}(\beta,s)} \leq \epsilon.
\]
It can be verified that a scaling of $s \sim \beta \epsilon^{-\frac{1}{2}}$ ($s\sim\epsilon^{-\frac{1}{2}}$) can satisfy this relative error for TQS (adaptive TQS). In fact, following the same technique of relative error of partition function, the same scaling of squeezing factor with $\beta$ and $\epsilon$ can be obtained.	
}
\end{document}